# Ordering effect on the electrical properties of stoichiometric $Ba_3CaNb_2O_9$-based perovskite ceramics

**Authors**

J. E. Rodrigues [1]; D. M. Bezerra [2]; A. C. Hernandes [1]

**Affiliations**

[1] São Carlos Institute of Physics, University of São Paulo, CEP 13560-970, São Carlos SP, Brazil.

[2] São Carlos Institute of Chemistry, University of São Paulo, CEP 13566-590, São Carlos SP, Brazil.

**Corresponding author**

J. E. Rodrigues

E-mail address: rodrigues.joaoelias@gmail.com or rodrigues.joaoelias@ursa.ifsc.usp.br.

URL: http://www.ifsc.usp.br/ccmc or http://cdmf.org.br.

## Abstract

Cation ordering is most common process detected in $A_3B'B''_2O_9$-based complex perovskites. Some important physical features of this system are due to the B-site ordering at long and short range. For microwave applications as filters and resonators, the 1:2 order is more appropriate. Otherwise, the oxygen vacancies and 1:1 order are considered the main factors behind the good performance of nonstoichiometric $A_3B'_{1+x}B''_{2-x}O_{9-\delta}$-based ceramics as proton conductors. Until now, however, there are no available reports regarding the isolated effects of B-site order at long range on the electrical properties of stoichiometric systems. This work reports the preparation of 1:1 and 1:2 fully-ordered $Ba_3CaNb_2O_9$ ceramics. Here, we combine the Raman scattering and group-theory calculations to distinguish the fingerprints of the 1:1 and 1:2 orders. The electric properties of the ordered $Ba_3CaNb_2O_9$ are analyzed in terms of a phenomenological model based on a parallel combination of a resistor, constant phase element, and capacitor. In particular, the conductivity relaxation ascribed to the grain is due to the oxygen vacancies. Besides, we found that the 1:1 order increases the *dc* conductivity, but not enough to account the good performance reported for the non-stoichiometric $Ba_3Ca_{1+x}Nb_{2-x}O_{9-\delta}$-based proton conductors.

## 1. Introduction

In the past decades, compounds based on ordered complex perovskites with the chemical formula $A_3B'B''_2O_9$ [A = $Ba^{2+}$ or $Sr^{2+}$; B' = $Mg^{2+}$, $Ca^{2+}$ or $Zn^{2+}$; B'' = $Nb^{5+}$ or $Ta^{5+}$] have attracted considerable attention owing to their very low loss and high proton conductivity, making them suitable for microwave devices [e.g. filters, resonators, antennas] [1], and electrolyte membranes for proton conducting solid oxide fuel cells [2]. In the former case, three parameters should be taken into account to improve the device performance: high dielectric permittivity enabling the miniaturization, low loss for good selectivity, and near zero temperature coefficient of permittivity providing a thermally stable device [3,4]. In the latter case, high proton conductivity and good chemical stability are desired capabilities to design new hydrogen and humidity sensors [5,6], chemical reactors for reforming of methane [7] and ammonia generation [8], and fuel cells [9]. In the perovskite structure, the protonic defects are usually due to the dissociative adsorption of water and/or hydrogen [10], in agreement with the next reactions [in Kröger-Vink notation]:

$$H_2O + V_O^{\bullet\bullet} + O_O^{\times} \leftrightarrow 2(OH)_O^{\bullet}, \quad (1)$$

$$H_2(g) + 2O_O^{\times} \leftrightarrow 2(OH)_O^{\bullet} + 2e', \quad (2)$$

where the protons are attached to oxide ions forming the $OH^-$ groups, with defect notation $(OH)_O^{\bullet}$ [11].

Particularly, the nonstoichiometric perovskite-structured $Ba_3Ca_{1+x}Nb_{2-x}O_{9-\delta}$ exhibited an appropriate chemical stability under water and $CO_2$ atmospheres at high temperatures for $x$ = 0.18 (BCN18) [12]. Nevertheless, its proton conductivity is low compared to the well-studied simple perovskites containing rare-earth-doped alkaline earth based cerates and zirconates [13–15]. Some strategies to enhance the proton conductivity of BCN18 without deteriorate its chemical stability have been reported, mainly focusing the generation of oxygen vacancies [$\equiv V_O^{\bullet\bullet}$]. The Thangadurai group has studied the partial replacement of Ba by K on A-site and Ti substitution for Nb in the stoichiometric $Ba_3CaNb_2O_9$ [16,17]; The Fanglin Chen group has reported the effects of Ce and Y doping on the Ca and Nb ions in complex perovskite $Ba_3Ca_{1.18}Nb_{1.82}O_{9-\delta}$ proton conductor [9,18]. In all cases, the nonstoichiometry leads to a formation of oxygen vacancies and a modification in the B-site ordering from 1:2 to 1:1 type [19]. In the 1:2 ordered systems, the Ca and Nb cations are alternately distributed in the sequence –Ca–Nb–Nb–Ca–Nb–Nb– along the $[111]_c$ direction of the parent cubic cell [20]. In the 1:1 order, the 4*b* sites are occupied by Nb, whereas the 4*a* ones possess a randomly distribution of Ca and Nb: ⅔Ca + ⅓Nb [21]. For microwave uses, the obtaining of low loss $A_3B'B''_2O_9$-based dielectrics is related to the 1:2 order [22–24]. To the best of our knowledge, however, there are no available reports regarding the isolated effects of B-site ordering on the electrical properties of stoichiometric $Ba_3CaNb_2O_9$-based ceramics.

In this paper, we synthesized the 1:1 and 1:2 ordered $Ba_3CaNb_2O_9$-based ceramics to probe the role of B-site ordering on their electrical properties, seeking to elucidate the ordering contribution to the *ac* conductivity and dielectric permittivity. As reported by many authors [25–27], the combination of both the complex impedance and complex electric modulus formalisms is an important tool to probe the conductivity relaxation in several materials, including conducting glasses, solid electrolytes, electroceramics, and so on. In particular, the electric modulus [$\mathbf{M}(\omega) \equiv 1/\boldsymbol{\varepsilon}^*(\omega) = \mathrm{Re}(\mathbf{M}) + j\mathrm{Im}(\mathbf{M})$] minimizes the occurrence of low frequency phenomena, such as the electrode polarization, parasite capacitance, and double layer formation, making such a formalism more appropriate to study the bulk contribution to the electrical response [28]. Besides, the signatures of order-disorder phenomena in ordered complex perovskites have been investigated using X-ray diffraction and Raman spectroscopy, mainly focusing on the fingerprints of short and long range ordering at B-site [29–31]. Such techniques will be used here to identify the ordering and then its role on the electrical properties of stoichiometric $Ba_3CaNb_2O_9$-based ceramics.

## 2. Experimental Procedures

### *2.1. Preparation of samples*

$Ba_3CaNb_2O_9$ [here after: BCNO] ceramics were prepared by solid-state reaction route under air atmosphere using a stoichiometric mixture of high purity $BaCO_3$ [Alfa Aesar; 99.80%], $CaCO_3$ [Alfa Aesar; 99.95%], and $Nb_2O_5$ [Alfa Aesar; 99.90%]. Such precursors were homogenized in a nylon jar containing isopropyl alcohol and zirconia cylinders during 24 h. The slurry was dried and subsequently calcined at 1573 K for 2 h. After milling, the BCNO powder was mixed with a binder solution of PVB-polyvinyl butyral [~ 3 wt%], pressed into cylindrical pellets [D ~ 6 mm × H ~ 2 mm] by uniaxial [15 MPa] and isostatic [300 MPa] cold pressing. To obtain 1:1 and 1:2 ordered ceramics, such green pellets were sintered in air by following the conditions listed in Table 1, yielding to dense ceramics with at least ~ 92% of relative density.

**Insert here: TABLE 1.**

### *2.2. Characterization of samples*

Powder X-ray diffraction patterns [pXRD] were recorded by a *Rigaku* Rotaflex RU-200B diffractometer [50 kV and 100 mA; Cu-Kα radiation λ = 1.5406 Å; Bragg-Brentano θ-2θ geometry] over a 2θ range from 10° to 100° with a step size of 0.02° and step time of 1 s. Raman spectra at room temperature were recorded using a *Jobin-Yvon*

T64000 triple monochromator with an $LN_2$ cooled CCD detector. A backscattering geometry was employed with a 488 nm line of an argon laser [Coherent INNOVA 70C] as an excitation source. The 1800 grooves.mm$^{-1}$ grating was chosen to keep the resolution below 1 cm$^{-1}$. Raman spectra were further corrected by the Bose-Einstein thermal factor prior to the fitting by the Lorentzian profiles [32], including a baseline subtraction. The microstructures were probed by an Inspect F50 field-emission scanning electron microscope [FEI, Netherlands] operating at 5 kV equipped with an energy dispersive X-ray spectrometer (EDX). The Archimedes principle was taken into account to obtain the bulk density of the ceramic samples using distilled water as an immersion liquid.

Measurements of impedance spectroscopy (IS) were performed in the frequency range from 0.5 Hz to 1 MHz using a Solartron frequency response analyzer [FRA model 1260] coupled with a dielectric interface Solartron [model 1296]. For this purpose, the pellet surfaces were painted with platinum paste (Pt), and then fired at 973 K for 1 h to remove its polymeric component and get good quality ceramic/electrode contact. The *ac* voltage of ~ 1V was defined after signal/noise tests, and considering the ohmic behavior of the system. The temperature dependence of the electrical properties of BCNO ceramics was acquired using a homemade furnace in the range from ~ 300 to 840 K under air atmosphere. Firstly, the thermal spectra of the permittivity [at 10 kHz, 100 kHz, and 1 MHz] were obtained in this temperature range, during the controlled cooling with rate of 2 K.min$^{-1}$. Next, the complex impedance spectra [$\mathbf{Z}^* = \mathrm{Re}(\mathbf{Z}) + j\mathrm{Im}(\mathbf{Z}) \equiv \lambda\boldsymbol{\rho}$; $\lambda$ is the geometrical factor in units of mm$^{-1}$] were acquired in the range between ~ 660 and 840 K, taking temperature plateaus with dwell time around 30 min in order to ensure a stable temperature (or isotherm) during the frequency sweep. The spectral data were further analyzed and adjusted using the ZView software 2.9c [33].

## 3. Results and Discussion

### *3.1. Cation ordering at B-site*

In this paper, we adjusted the synthesis conditions for obtaining fully disordered, 1:1 and 1:2 ordered BCNO samples. The disordered state takes place during the powder sample preparation, where $Ca^{2+}$ and $Nb^{5+}$ cations are randomly distributed at the B-site in $Ba(Ca,Nb)O_3$ composition [34]. Such a crystal structure belongs to the cubic space group *Pm*-3*m* [lattice constant: $a_0$] without octahedral tilting [$\boldsymbol{a}^0\boldsymbol{a}^0\boldsymbol{a}^0$ in Glazer notation] [35], and its pXRD pattern is illustrated in Fig. 1 (a). The ordered state can be achieved at high temperature through the sintering process [36], see details in Table 1. By monitoring the sintering, the 1:1 and 1:2 ordering types were obtained for the BCNO ceramic samples. In both cases, a superstructure is formed from the disordered cubic cell, leading to an increasing in the unit cell volume [37]. The signature of this phenomenon can be detected in Fig. 1 (b) through the superlattice X-ray Bragg reflections. Since the sizes and valences of $Ca^{2+}$ and $Nb^{5+}$ are found to be different, their Wyckoff sites should contain distinct local properties depending on the Ca–O and Nb–O bond distances. This fact explains the appearing of the superlattice peaks at lower diffraction angles, labeled as $(111)_\mathbf{c}$, $(001)_\mathbf{h}$ and $(100)_\mathbf{h}$ [24,38]. The detection of those superlattice reflections is an evidence for the long range 1:1 and 1:2 orderings in the BCNO ceramics [39].

**Insert here: FIGURE 1.**

In the $Ba_3CaNb_2O_9$-like system, the $(001)_\mathbf{h}$ and $(100)_\mathbf{h}$ superlattice peaks come from the –Ca–Nb–Nb–Ca–Nb–Nb– stacking sequence (1:2) along the $[111]_\mathbf{c}$ planes of the simple cubic cell [disordered system P*m*-3*m*], leading to a trigonal cell with the lattice parameters $a_\mathbf{h} \approx \sqrt{2}a_0$ and $c_\mathbf{h} \approx \sqrt{3}a_0$ [space group: *P*-3*m*1 or $D_{3d}^3$] [38]. However, such a composition may also be indexed based on the face-centered cubic structure with cell dimension $a_F \approx 2a_0$ [space group: $Fm$-3$m$ or $O_h^5$], leading to the $(111)_\mathbf{c}$ superlattice Bragg reflection [21]. In this case, the stacking sequence (1:1) has the following rule [40]: –(⅔Ca + ⅓Nb)–Nb–(⅔Ca + ⅓Nb)–Nb–. It means that there is at least one crystallographic site with a random cation distribution (4$a$ site): the so-called random site model [41]. Fig. 1 (c) displays a peak splitting process, namely $(110)_\mathbf{c} \rightarrow (220)_\mathbf{c} \rightarrow (012)_\mathbf{h} + (110)_\mathbf{h}$, owing to the trigonal cell formation. Therefore, at this point, it can be concluded that we have prepared two ceramic samples with well-defined long range 1:1 and 1:2 orderings, respectively. Then, one can expect different physical properties coming from these two ordering types. In order to probe how the structural ordering at short and long range impacts the Raman spectra of BCNO samples at room temperature, the group theory analysis for the crystal structures addressed here are reported in Table 2. In this theoretical approach, the number and symmetries of the optical active modes can be estimated, based on the occupied Wyckoff sites within the crystalline structure [42].

**Insert here: TABLE 2.**

For the fully disordered $Ba(Ca,Nb)O_3$ structure, no Raman-active modes should be observed in the first order Raman spectrum. Otherwise, 4 [$2\mathbf{F}_{2g} \oplus \mathbf{E}_g \oplus \mathbf{A}_{1g}$] and 9 [$5\mathbf{E}_g \oplus 4\mathbf{A}_{1g}$] Raman-active modes are expected to appear in the first order Raman spectra of 1:1 and 1:2 ordered samples, respectively. Fig. 2 (a) displays the Raman spectra acquired at room temperature for the disordered [powder], 1:1 and 1:2 ordered [ceramics] samples. For the 1:2 ceramic, the number of detected peaks agrees well with those predicted by the factor-group analysis. Particularly, the bands at 135, 250, 280 and 610 $cm^{-1}$ are fingerprints of the trigonal unit cell [38,43]. The internal modes of $NbO_6$ octahedra are located at 355, 410 and 818 $cm^{-1}$. The low wavenumber peaks at 85 and 90 $cm^{-1}$ are called the external modes concerning the vibrations of $Ba^{2+}$ ions against $O^{2-}$ anions. For the 1:1 ceramic, the broad bands at ~ 90, 300, 550 and 800 $cm^{-1}$ are assigned to $\mathbf{F}_{2g}$, $\mathbf{F}_{2g}$, $\mathbf{E}_{g}$, and $\mathbf{A}_{1g}$ modes [44]. Besides, the last mode concerns the breathing vibration of the oxygen octahedra. For the disordered sample, however, its Raman spectrum containing at least 11 bands more closely resembles that of the 1:2 ceramic. It means that the disordered sample has a 1:2 order at short range, since the superlattice reflections did not appear in its pXRD pattern in Fig. 1 (b). As accepted in recent literature, in this situation there are domain regions with 1:2 order embedded in a disordered cubic matrix [average space group: *Pm*-3*m* or $O_h^1$] [45]. Similar model was taken into account to assign the first order Raman spectra of $Pb_3MgNb_2O_9$-based relaxor ferroelectrics, but considering the domain regions with 1:1 order instead of 1:2 order [46,47]. However, further work should be carried out in order to account the short range features in BCNO.

**<u>Insert here: FIGURE 2.</u>**

It is worth mentioning that the $\mathbf{A}_{1g}$ mode in the wavenumber range 700–900 $cm^{-1}$ is very sensitive to the cation ordering at B-site [30,34,48]. Such a mode denotes a breathing type vibration of oxygen atoms, being its wavenumber dependent on the B–O and O–O bond strengths [49]. Therefore, the $\mathbf{A}_{1g}$ mode properties such as wavenumber, width and intensity should reflect the B-site order changes [50]. Fig. 2 (b) shows the $\mathbf{A}_{1g}$ peak fitting procedure for the disordered, 1:1 and 1:2 ordered samples. It can be noticed a strongest peak centered at 818 $cm^{-1}$ with a weak shoulder at 750 $cm^{-1}$ in the 1:2 sample spectrum. Otherwise, such a shoulder becomes more intense and shifts to higher wavenumber (at 765 $cm^{-1}$) for the 1:1 sample. For the disordered one, the band at 773 $cm^{-1}$ behaves like a shoulder again. Since we have checked the long range order in the 1:1 and 1:2 ceramic samples, it can be further concluded that the $\mathbf{A}_{1g}$ modes at ~ 760 and 820 $cm^{-1}$ are fingerprints of the 1:1 and 1:2 cation ordering in the $Ba_3CaNb_2O_9$ system, respectively. Fig. 2 (c) displays the dependence of the perceptual area of 1:1 and 1:2 peaks on the order state. In this sense, the 1:2 ceramic should contain large 1:2 or-dered domains and 1:1 domain boundary, as recently demonstrated by Ma *et al* for the $Ba_3(Co, Zn, Mg)Nb_2O_9$ perovskites [51]. On the contrary, the existence of large 1:1 ordered domain should better describe the Raman spectrum of 1:1 ceramic, with some regions containing short range 1:2 order, since we have detected a mode at ~ 827 $cm^{-1}$.

*3.2. Microstructural properties*

To probe the microstructural features of the 1:1 and 1:2 ceramics, we performed an elemental mapping by EDX analysis on the polished and thermally etched surfaces of the ceramic samples, as shown in Fig. 3 (a-h). As one can note, the Ba L, Ca K, Nb L and O K elements are homogeneously distributed in the SEM micrographs. Such a set of images allows concluding that there is no inter-granular phase associated with possible segregation at grain boundaries around triple junctions. Regarding the grain size distribution by means of the log-normal profile, its average value varied from ~ 2.8 to 1.4 μm for the 1:1 and 1:2 ceramics, respectively. All grain size distributions were performed using at least ~ 300 grains per ceramic. In a recent work, we have demonstrated in the BCNO system that the dielectric loss at microwave frequency can be tuned by controlling the coarsening of the 1:2 ordered domains [24]. Next, the impedance spectroscopy is applied to probe the role of the B-site cation ordering in the *ac* conductivity and permittivity of the $Ba_3CaNb_2O_9$-based perovskite ceramics. Here, we are interested in exploring the effect of order state in the electrical properties of BCNO dense ceramics, and particularly the contributions ascribed to the grains [or bulk].

**Insert here: FIGURE 3.**

*3.3. Electrical conductivity and dielectric properties*

For a comparative analysis, Fig. 4 shows (a) the thermal spectra of the permittivity, derived from the 10 kHz, 100 kHz and 1 MHz capacitance data, and (b) the Nyquist plot [or Cole-Cole: -Im(**ρ**) vs. Re(**ρ**)] at ~ 723 K for BCNO ceramics. The permittivity values are found to be different for the 1:1 and 1:2 ordered ceramics, and decreased with increasing temperature. It should be noticed that only one well-resolved semicircle for each sample appeared in the Nyquist plot of Fig. 4 (b). Fig. 4 (c) and Fig. 4 (d) illustrate the resistivity spectra of the 1:1 and 1:2 ceramics at selected temperatures, respectively. There is no evidence of a second contribution ascribed to the grain boundary and/or electrode polarization. Such effects are usually detected as spikes in low frequency part of the Nyquist plot [52]. In order to describe the complex resistivity of our samples, we used a phenomenological model based on a parallel combination of a resistor ($R_g$), constant phase element (CPE), and capacitor ($C_g$), see in Fig. 5. Here, the $R_S$ element takes into account the resistance coming from the measuring system. Hernández, Masó, and West have proposed this circuit model to improve the fitting procedure of the conductivity spectra of solid electrolytes and electroceramics [ionic and electronic conductors] [53]. In practice, such a model introduces a limiting high frequency value ($\varepsilon_g$) in the real permittivity spectra. From Fig. 4, there is a good agreement between experimental data and simulated curves.

**Insert here: FIGURE 4, and FIGURE 5.**

Table 3 summarizes a comparison among three circuit models, namely: $R_g||C_g$, $R_g||CPE$, and $R_g||CPE||C_g$. For parallel combination of a resistor and capacitor ($R_g||C_g$), a non-dispersive behavior is expected, since Re(**σ**) = $\sigma_g$ and Re(**ε**) = $\varepsilon_g$ [26]. Otherwise, the electrical response of real systems is better described by considering their dispersive features, and particularly Re($\boldsymbol{\sigma}$) ~ $\omega^{\alpha}$ and Re($\boldsymbol{\varepsilon}$) ~ $\omega^{\alpha-1}$ [54]. In both models containing CPE, the real conductivity exhibits the same dependency. Otherwise, the real permittivity provides a limiting value ($\varepsilon_g$) at high frequencies only for the $R_g$||CPE||$C_g$ model. Owing to its accuracy we have employed this model to obtain valuable information on the electrical response of the ceramic samples addressed in this work. It is worth mentioning that the CPE introduces the dispersion behavior to take into account the Jonscher's Universal Dielectric Response (UDR) [55,56]. At each temperature, the real and imaginary parts of the complex resistivity were fitted by the $R_g$||CPE||$C_g$ model, as depicted in Fig. 6 (a) and Fig. 6 (c) for the 1:1 and 1:2 ceramic samples, respectively. Regarding the CPE, two parameters should be adjusted, and particularly its pseudo-capacitance Q and the dimensionless factor α (0 < α < 1) [57]. The CPE impedance is expressed as follows:

$$Z_{CPE} = \frac{1}{Q(j\omega)^{\alpha}}. \quad (3)$$

**Insert here: FIGURE 6, and TABLE 3.**

The peak in the imaginary part of resistivity indicates a relaxation frequency associated to a conductivity relaxation time $\tau_\sigma$, in which $\omega_m\tau_\sigma = 1$. Such a process can be also seen in the imaginary part of electric modulus [$\mathbf{M}(\omega)$] of Fig. 6 (b) and Fig. 6 (d). Both peaks are found to be almost symmetrical in the frequency range from 1 Hz to $10^4$ Hz. Furthermore, the relaxation frequencies are temperature dependent, and defining a thermally-activated process. From Fig. 7, it can be observed such an effect more clearly through the contour map of the imaginary part of electric modulus as a function of temperature. Indeed, the conductivity relaxation frequency ($f_m = \omega_m/2\pi$) increases with increasing temperature for both BCNO ceramics. Fig. 8 (a) and Fig. 8 (c) represent the dispersion behavior (~ $\omega^{\alpha}$) of the real conductivity spectra at selected temperatures. As mentioned earlier, each spectrum contains a low frequency plateau ascribed to the *dc* conductivity $\sigma_g$ (= $\lambda/R_g$). These values are found to be different for the 1:1 and 1:2 ceramic samples. Such quantities are also sensitive to changes in temperature, being defined as a thermally-activated process [58]. In order to analyze the temperature dependence of the *dc* conductivity and relaxation frequency ($f_m$), we have plotted in Fig. 8 (b) and Fig. 8 (d) the reciprocal temperature dependence of those quantities for the 1:1 and 1:2 ceramics, respectively. In both cases, the general trend of the curves suggested an Arrhenius-like behavior, as follows:

$$\sigma_g T = A_\sigma \exp\left(-\frac{E_\sigma}{kT}\right), \quad (4)$$

$$f_m T = B_f \exp\left(-\frac{E_f}{kT}\right). \quad (5)$$

From these equations, one can find the macroscopic activation energies: ~ 1.6 eV [1:1] and 1. 7 eV [1:2]. It should be noticed that the activation energies derived from *dc* conductivity and relaxation frequency exhibited almost the same value [$E_\sigma \approx E_f$], denoting an ionic-hopping mechanism [here oxygen vacancies: $V_O^{\bullet\bullet}$] to the electrical conduction in the 1:1 and 1:2 ceramics [59]. Indeed, the magnitude of the activation energy determined here agrees well with those reported for ionic conductor ceramics [60,61].

Concerning the real part of dielectric permittivity, it was detected a low frequency dispersion (~ $\omega^{\alpha-1}$) followed by a plateau at high frequencies associated with the limiting value $\varepsilon_g$ (= $\lambda C_g$). Such a value is temperature dependent as we have demonstrated in Fig. 4 (a). In the temperature range between ~ 660 and 840 K, the temperature coefficient of permittivity [$\tau_\varepsilon$ in units of ppm.$K^{-1}$] was calculated, in accordance with the next equation:

$$\tau_\varepsilon = \frac{1}{\varepsilon_{g,i}} \frac{\Delta\varepsilon_g}{\Delta T}, \quad (6)$$

so that $\varepsilon_{g,i}$ is the high frequency permittivity at around 660 K. From Table 1, one should note different values for the temperature coefficient $\tau_\varepsilon$ depending on the long range order. Besides, the negative value for $\tau_\varepsilon$ denotes its decrease with increasing temperature. It should be highlighted a good agreement between the permittivity data at 100 kHz or 1 MHz and permittivity $\varepsilon_g$ values extracted from fitting complex resistivity spectra to the $R_g$||CPE||$C_g$ circuit model.

**Insert here: FIGURE 7, and FIGURE 8.**

From Table 1, it is reasonable to postulate that the difference in both *dc* conductivity and permittivity stems from the ordering type of the crystal structure. In light of the Brick-Layer model, extrinsic properties such as grain size and grain boundary thickness are only decisive to modulate the electrical response coming from the grain boundaries [62,63]. In this way, intrinsic factors should better describe the electrical properties of the grains, including the cation ordering. For the $A_3B'B''_2O_9$-like compounds, the 1:1 order yields a disorder at 4*a* site with a random distribution: (⅔Ca + ⅓Nb). In a qualitative view, this disorder may lead to an anharmonicity of the crystal lattice, and then increasing the dielectric loss (tan δ). The electrical conductivity is sensitivity to the dielectric loss changes, as predicted by the following identity:

$$\mathrm{Re}(\sigma) = \omega(\tan\delta)\mathrm{Re}(\varepsilon). \quad (7)$$

The earlier assumption is based on the results reported in the literature concerning the dependence of dielectric loss on disorder in $A_3B'B''_2O_9$ perovskites. In general, an increased dielectric loss is expected for such systems containing a local disorder at B-site.

Here, our results have illustrated that the disorder effects modify the conductivity relaxation process in the grains (or bulk) of the $Ba_3CaNb_2O_9$-based ceramics.

In the literature, BCNO systems have been investigated for possible applications as proton conductor for solid oxide fuel cells [9,18]. As we have demonstrated, the 1:2 ordered $Ba_3CaNb_2O_9$ depicts an insulator behavior at temperatures below ~ 673 K [$\sigma_g < 10^{-10}$ $(\Omega.mm)^{-1}$]. Concerning the 1:1 ordered BCNO, its *dc* conductivity is always higher than that of 1:2 ceramic, but not reaching the upper values reported for the $Ba_3Ca_{1+x}Nb_{2-x}O_{9-\delta}$ system. In this material, two processes take place with the substitution of $Ca^{2+}$ for $Nb^{5+}$: oxygen vacancy generation and gradual transition from 1:2 trigonal to 1:1 cubic [19,64]. From the defect chemistry point of view, the formation of oxygen vacancies is designated by the next reaction [in Kröger-Vink notation]:

$$3O_O^{\times} + 2Nb_{Nb}^{\times} + 2CaO \rightarrow 3V_O^{\bullet\bullet} + 2Ca_{Nb}^{\prime\prime\prime} + Nb_2O_5 . \quad (8)$$

Based on atomistic simulations, Ruiz-Trejo *et al* have concluded that the oxygen migration along $NbO_6$ octahedra is most preferred in comparison with $CaO_6$ octahedra, since the migration energies [~ 1 eV] for oxygen jumps along the edges of $NbO_6$ octahedra are less than those energies [~ 2 eV] for jumps along the edges of $CaO_6$ units [65].

Interestingly, the reported values between 0.70 and 0.92 eV suggests that the oxygen migration along $NbO_6$ occurs in the $Ba_3Ca_{1+x}Nb_{2-x}O_{9-\delta}$ ceramics [66]. In this system, a percolation path of $NbO_6$ octahedra accounts the increase in the proton conductivity until $x = 0.18$. For $x > 0.18$, it is observed an increased activation energy, and then a decrease in the *dc* conductivity despite the high concentration of oxygen vacancies [17,64]. From our results above, we can conclude that the 1:1 ordering enhances the *dc* conductivity, but not enough to account the excellent performance reported for the non-stoichiometric barium calcium niobate ceramics. Based on the macroscopic activation energies between 1.62 [1:1] and 1. 71 eV [1:2], a more complex oxygen migration along both $CaO_6$ and $NbO_6$ units should take place in the 1:1 and 1:2 ordered BCNO ceramics investigated in the present work.

## 4. Conclusions

In short, we have investigated the 1:1 and 1:2 ordered $Ba_3CaNb_2O_9$-based perovskite ceramics prepared by conventional solid state method. Our main goal in this paper was to probe the role of B-site ordering at long range on the electrical properties of stoichiometric $Ba_3CaNb_2O_9$ ceramics. The synthesis conditions were adjusted for obtaining fully 1:1 and 1:2 ordered BCNO. The ordered structures could be detected by the occurrence of superlattice X-ray reflections at lower 2θ angles, being an evidence for the long range order. A combination of Raman scattering and factor-group analysis was promising to detect the fingerprints of the 1:1 and 1:2 ordering through the $\mathbf{A}_{1g}$ modes at ~ 760 and 820 $cm^{-1}$. We have used a circuit model based on a parallel combination of a resistor, constant phase element, and capacitor to describe the complex resistivity spectra of our ceramics. The conductivity relaxation coming from the grain is ascribed to the oxygen vacancies. From our results here, we have argued that the 1:1 order enhances the *dc* conductivity, but not enough to account the excellent performance reported for the non-stoichiometric $Ba_3Ca_{1+x}Nb_{2-x}O_{9-\delta}$.

## Acknowledgments

The authors are grateful to the financial support of the Brazilian agencies CAPES [BEX 0270/16-4], CNPq [Proc. N.° 573636/2008-7], INCTMN [Proc. N.° 2008/57872-1], and FAPESP [Proc. N.° 2013/07296- 2]. J. E. would like to thank Pr. Olivier Thomas for all support at Aix-Marseille Université.

## References

[1] I.M. Reaney, D. Iddles, Microwave dielectric ceramics for resonators and filters in mobile phone networks, J. Am. Ceram. Soc. 89 (2006) 2063–2072.

[2] E. Fabbri, D. Pergolesi, E. Traversa, Materials challenges toward proton-conducting oxide fuel cells: a critical review., Chem. Soc. Rev. 39 (2010) 4355–69.

[3] T.A. Vanderah, Materials science. Talking ceramics., Science. 298 (2002) 1182–1184.

[4] M. Sebastian, Dielectric Materials for Wireless Communication, 1st ed, Elsevier, 2008.

[5] H. Matsumoto, T. Suzuki, H. Iwahara, Automatic regulation of hydrogen partial pressure using a proton conducting ceramic based on SrCeO3, Solid State Ionics. 116 (1999) 99–104.

[6] M.. Hassen, A.. Clarke, M.. Swetnam, R.. Kumar, D.. Fray, High temperature humidity monitoring using doped strontium cerate sensors, Sensors Actuators B Chem. 69 (2000) 138–143.

[7] T. Hibino, S. Hamakawa, T. Suzuki, H. Iwahara, Recycling of carbon dioxide using a proton conductor as a solid electrolyte, J. Appl. Electrochem. 24 (1994) 126–130.

[8] G. Marnellos, M. Stoukides, Ammonia Synthesis at Atmospheric Pressure, Science (80-. ). 282 (1998) 98–100.

[9] S. Wang, Y. Chen, S. Fang, L. Zhang, M. Tang, K. An, et al., Novel Chemically Stable Ba 3 Ca 1.18 Nb 1.82− x Y x O 9−δ Proton Conductor: Improved Proton

Conductivity through Tailored Cation Ordering, Chem. Mater. 26 (2014) 2021–2029.

[10] I. Kosacki, H. Tuller, Mixed conductivity in $SrCe_{0.95}Yb_{0.05}O_3$ protonic conductors, Solid State Ionics. 80 (1995) 223–229.

[11] T. Norby, Proton conduction in oxides, Solid State Ionics. 40–41 (1990) 857–862.

[12] D. Hassan, S. Janes, R. Clasen, Proton-conducting ceramics as electrode/electrolyte materials for SOFC's—part I: preparation, mechanical and thermal properties of sintered bodies, J. Eur. Ceram. Soc. 23 (2003) 221–228.

[13] J. Wu, R.A. Davies, M.S. Islam, S.M. Haile, Atomistic Study of Doped BaCeO 3 : Dopant Site-Selectivity and Cation Nonstoichiometry, Chem. Mater. 17 (2005) 846–851.

[14] S. Wang, F. Zhao, L. Zhang, F. Chen, Synthesis of $BaCe_{0.7}Zr_{0.1}Y_{0.1}Yb_{0.1}O_{3-\delta}$ proton conducting ceramic by a modified Pechini method, Solid State Ionics. 213 (2012) 29–35.

[15] R. Kannan, K. Singh, S. Gill, T. Fürstenhaupt, V. Thangadurai, Chemically Stable Proton Conducting Doped $BaCeO_3$ -No More Fear to SOFC Wastes, Sci. Rep. 3 (2013) 2138.

[16] S.S. Bhella, V. Thangadurai, Investigations on the thermo-chemical stability and electrical conductivity of K-doped $Ba_{3-x}K_xCaNb_2O_{9-\delta}$ (x=0.5, 0.75, 1, 1.25), Solid State Ionics. 192 (2011) 229–234.

[17] T.T. Trinh, V. Thangadurai, Effect of Ti substitution for Nb in double perovskite-type $Ba_3CaNb_2O_9$ on chemical stability and electrical conductivity,

Electrochim. Acta. 56 (2010) 227–234.

[18] S. Wang, F. Zhao, L. Zhang, K. Brinkman, F. Chen, Doping effects on complex perovskite Ba3Ca1.18Nb1.82O9−δ intermediate temperature proton conductor, J. Power Sources. 196 (2011) 7917–7923.

[19] O. Valdez-Ramírez, F. Gómez-García, M.A. Camacho-López, E. Ruiz-Trejo, Influence of the calcium excess in the structural and spectroscopic properties of the complex perovskite Ba3CaNb2O9, J. Electroceramics. 28 (2012) 226–232.

[20] I.M. Reaney, I. Qazi, W.E. Lee, Order–disorder behavior in Ba(Zn1/3Ta2/3)O3, J. Appl. Phys. 88 (2000) 6708–6714.

[21] W.H. Kan, J. Lussier, M. Bieringer, V. Thangadurai, Studies on polymorphic sequence during the formation of the 1:1 ordered perovskite-type BaCa(0.335)M(0.165)Nb(0.5)O(3-δ) (M = Mn, Fe, Co) using in situ and ex situ powder X-ray diffraction., Inorg. Chem. 53 (2014) 10085–93.

[22] D.A. Sagala, S. Nambu, Microscopic Calculation of Dielectric Loss at Microwave Frequencies for Complex Perovskite Ba(Zn1/3Ta2/3)O3, J. Am. Ceram. Soc. 75 (1992) 2573–2575.

[23] A. Dias, R.L. Moreira, Far-infrared spectroscopy in ordered and disordered BaMg[sub 1/3]Nb[sub 2/3]O[sub 3] microwave ceramics, J. Appl. Phys. 94 (2003) 3414.

[24] J.E.F.S. Rodrigues, P.J. Castro, P.S. Pizani, W.R. Correr, A.C. Hernandes, Structural ordering and dielectric properties of Ba3CaNb2O9-based microwave ceramics, Ceram. Int. 42 (2016) 18087–18093.

[25] I.M. Hodge, M.D. Ingram, A.R. West, Impedance and modulus spectroscopy of

polycrystalline solid electrolytes, J. Electroanal. Chem. Interfacial Electrochem. 74 (1976) 125–143.

[26] J. Świergiel, J. Jadżyn, Electric Relaxational Effects Induced by Ionic Conductivity in Dielectric Materials, Ind. Eng. Chem. Res. 50 (2011) 11935–11941.

[27] J.E.F.S. Rodrigues, C.W.D.A. Paschoal, E.N. Silva, K.A. Mince, M.W. Lufaso, Relaxations in Ba2BiTaO6 ceramics investigated by impedance and electric modulus spectroscopies, Mater. Res. Bull. 47 (2012) 878–882.

[28] E. Barsoukov, J.R. Macdonald, Impedance Spectroscopy: Theory, Experiment, and Applications, 2nd ed, Wiley, 2005.

[29] C.-H. Lu, C.-C. Tsai, Reaction kinetics, sintering characteristics, and ordering behavior of microwave dielectrics: Barium magnesium tantalate, J. Mater. Res. 11 (1996) 1219–1227.

[30] R.L. Moreira, F.M. Matinaga, A. Dias, Raman-spectroscopic evaluation of the long-range order in Ba(B'1/3B''2/3)O3 ceramics, Appl. Phys. Lett. 78 (2001) 428.

[31] C.-H. Wang, X.-P. Jing, L. Wang, J. Lu, XRD and Raman Studies on the Ordering/Disordering of Ba(Mg 1/3 Ta 2/3 )O 3, J. Am. Ceram. Soc. 92 (2009) 1547–1551.

[32] W. Hayes, R. Loudon, Scattering of Light by Crystals, Dover, New York, 2004.

[33] D. Johnson, Software ZView 3.2, Southern Pines: Scribner Associates, Inc, 2009.

[34] I.G. Siny, R.S. Katiyar, A.S. Bhalla, Cation arrangement in the complex

perovskites and vibrational spectra, J. Raman Spectrosc. 29 (1998) 385–390.

[35] F.S. Galasso, R. Smoluchowski, N. Kurti, Structure, Properties and Preparation of Perovskite-Type Compounds, Elsevier Science, 1969.

[36] B. Burton, Empirical cluster expansion models of cation order-disorder in A(B1/3′,B2/3″)O3 perovskites, Phys. Rev. B. 59 (1999) 6087–6091.

[37] D.A. Sagala, S. Nambu, Lattice Energy Calculations for Ordered and Disordered B a ( Z n 1/3 T a 2/3 ) O 3, J. Phys. Soc. Japan. 61 (1992) 1791–1797.

[38] J. Deng, J. Chen, R. Yu, G. Liu, X. Xing, Crystallographic and Raman spectroscopic studies of microwave dielectric ceramics Ba(Ca1/3Nb2/3)O3, J. Alloys Compd. 472 (2009) 502–506.

[39] A. Dias, V.S.. Ciminelli, F.. Matinaga, R.. Moreira, Raman scattering and X-ray diffraction investigations on hydrothermal barium magnesium niobate ceramics, J. Eur. Ceram. Soc. 21 (2001) 2739–2744.

[40] J. Chen, H.M. Chan, M.P. Harmer, Ordering Structure and Dielectric Properties of Undoped and La/Na-Doped Pb(Mg1/3Nb2/3)O3, J. Am. Ceram. Soc. 72 (1989) 593–598.

[41] L. Chai, M.A. Akbas, P.K. Davies, J.B. Parise, Cation ordering transformations in Ba(MgTa)O3-BaZrO3 perovskite solid solutions, Mater. Res. Bull. 32 (1997) 1261–1269.

[42] D.L. Rousseau, R.P. Bauman, S.P.S. Porto, Normal mode determination in crystals, J. Raman Spectrosc. 10 (1981) 253–290.

[43] J.E.F.S. Rodrigues, E. Moreira, D.M. Bezerra, A.P. Maciel, C.W. de Araujo Paschoal, Ordering and phonons in Ba3CaNb2O9 complex perovskite, Mater.

Res. Bull. 48 (2013) 3298–3303.

[44] C.-L. Diao, C.-H. Wang, N.-N. Luo, Z.-M. Qi, T. Shao, Y.-Y. Wang, et al., First-Principle Calculation and Assignment for Vibrational Spectra of Ba(Mg 1/2 W 1/2 )O 3 Microwave Dielectric Ceramic, J. Am. Ceram. Soc. 96 (2013) 2898–2905.

[45] F. Jiang, S. Kojima, C. Zhao, C. Feng, Chemical ordering in lanthanum-doped lead magnesium niobate relaxor ferroelectrics probed by A[sub 1g] Raman mode, Appl. Phys. Lett. 79 (2001) 3938.

[46] C.A. Randall, A.S. Bhalla, Nanostructural-Property Relations in Complex Lead Perovskites, Jpn. J. Appl. Phys. 29 (1990) 327–333.

[47] F. Jiang, S. Kojima, C. Zhao, C. Feng, Raman scattering on the B-site order controlled by A-site substitution in relaxor Perovskite ferroelectrics, J. Appl. Phys. 88 (2000) 3608–3612.

[48] M.Y. Chen, P.J. Chang, C.T. Chia, Y.C. Lee, I.N. Lin, L.-J. Lin, et al., Extended X-ray absorption fine structure, X-ray diffraction and Raman analysis of nickel-doped Ba(Mg1/3Ta2/3)O3, J. Eur. Ceram. Soc. 27 (2007) 2995–2999.

[49] N.S.N. Nath, The normal vibrations of molecules having octahedral symmetry, Proc. Indian Acad. Sci. - Sect. A. 1 (1934) 250–259.

[50] G. Blasse, A.F. Corsmit, Vibrational spectra of 1:2 ordered perovskites, J. Solid State Chem. 10 (1974) 39–45.

[51] P.P. Ma, H. Gu, X.M. Chen, Determination of 1:2 Ordered Domain Boundaries in Ba[(Co, Zn , Mg)1/3Nb2/3]O3 Dielectric Ceramics, J. Am. Ceram. Soc. 99 (2016) 1299–1304.

[52] R.V. Barde, S.A. Waghuley, Study of AC electrical properties of V2O5–P2O5–B2O3–Dy2O3 glasses, Ceram. Int. 39 (2013) 6303–6311.

[53] M.A. Hernández, N. Masó, A.R. West, On the correct choice of equivalent circuit for fitting bulk impedance data of ionic/electronic conductors, Appl. Phys. Lett. 108 (2016) 152901.

[54] D.P. Almond, A.R. West, Impedance and modulus spectroscopy of “real” dispersive conductors, Solid State Ionics. 11 (1983) 57–64.

[55] A.K. Jonscher, The “universal” dielectric response, Nature. 267 (1977) 673–679.

[56] A.K. Jonscher, Relaxation in low-loss dielectrics, J. Mol. Liq. 86 (2000) 259–268.

[57] M.R. Shoar Abouzari, F. Berkemeier, G. Schmitz, D. Wilmer, On the physical interpretation of constant phase elements, Solid State Ionics. 180 (2009) 922–927.

[58] B.T. Mark E. Orazem, Electrochemical Impedance Spectroscopy, Wiley, 2008.

[59] S. Lanfredi, P.S. Saia, R. Lebullenger, A.C. Hernandes, Electric conductivity and relaxation in fluoride, fluorophosphate and phosphate glasses: analysis by impedance spectroscopy, Solid State Ionics. 146 (2002) 329–339.

[60] X. Kuang, X. Jing, Z. Tang, Dielectric loss spectrum of ceramic MgTiO3 investigated by AC impedance and microwave resonator measurements, J. Am. Ceram. Soc. 89 (2006) 241–246.

[61] M. Malki, J. Schreuer, H. Schneider, Electrical conductivity of synthetic mullite single crystals, Am. Mineral. 99 (2014) 1104–1108.

[62] T. van Dijk, A.J. Burggraaf, Grain boundary effects on ionic conductivity in

ceramic GdxZr1–xO2–(x/2) solid solutions, Phys. Status Solidi. 63 (1981) 229–240.

[63] J. Fleig, J. Maier, The impedance of ceramics with highly resistive grain boundaries: validity and limits of the brick layer model, J. Eur. Ceram. Soc. 19 (1999) 693–696.

[64] Y. Du, A.S. Nowick, Structural Transitions and Proton Conduction in Nonstoichiometric A3B'B"O9 Perovskite-Type Oxides, J. Am. Ceram. Soc. 78 (1995) 3033–3039.

[65] E. Ruiz-Trejo, R.A. De Souza, Dopant substitution and oxygen migration in the complex perovskite oxide Ba3CaNb2O9: A computational study, J. Solid State Chem. 178 (2005) 1959–1967.

[66] A. Nowick, Y. Du, K. Liang, Some factors that determine proton conductivity in nonstoichiometric complex perovskites, Solid State Ionics. 125 (1999) 303–311.

## Figure Captions

**Figure 1:** (a) Powder X-Ray diffraction patterns of disordered, 1:1 and 1:2 ordered BCNO samples at room temperature; (b) superlattice X-Ray Bragg reflections at lower diffraction angles; (c) Peak splitting process, namely $(110)_{\mathbf{c}} \rightarrow (220)_{\mathbf{c}} \rightarrow (012)_{\mathbf{h}} + (110)_{\mathbf{h}}$, owing to the trigonal cell formation.

**Figure 2:** (a) Raman spectra at room temperature for the disordered, 1:1 and 1:2 ordered samples; (b) The black open circles denote experimental data and red/green lines are the fitted spectrum using Lorentzian profiles; (c) Dependence of the 1:1 and 1:2 peak areas with the order state.

**Figure 3:** EDX elemental mappings toward the SEM micrographs on the left of the 1:1 (a-d) and 1:2 (e-h) BCNO ceramics. There is no inter-granular phase associated with possible segregation at grain boundaries around triple junctions.

**Figure 4:** (a) Thermal spectra of the dielectric permittivity, derived from the 10 kHz, 100 kHz and 1 MHz capacitance data, and limiting value $\varepsilon_g$; (b) the Nyquist plot [-Im($\boldsymbol{\rho}$) vs. Re($\boldsymbol{\rho}$)] at ~ 723 K for the 1:1 and 1:2 ordered BCNO ceramics. The resistivity spectra of the 1:1 (c) and 1:2 (d) ordered ceramics at selected temperatures. There is no evidence of a second contribution coming from the grain boundary and/or electrode polarization.

**Figure 5:** A phenomenological model based on a parallel combination of resistor ($R_g$), constant phase element (CPE), and capacitor ($C_g$). The $R_S$ element denotes the resistance coming from the measuring system.

**Figure 6:** The real and imaginary parts of the complex resistivity of the 1:1 (a) and 1:2 (c) ordered ceramics. The imaginary parts of the electric modulus spectra of the 1:1 (b) and 1:2 (d) ceramics. There is a good agreement between experimental data and simu-

lated curves.

**Figure 7:** The contour maps of the imaginary part of electric modulus as a function of temperature for the 1:1 (a) and 1:2 (b) ordered BCNO ceramics. The relaxation frequencies are temperature dependent, and defining a thermally-activated process.

**Figure 8:** The dispersion behavior ($\sim \omega^{\alpha}$) of the real conductivity spectra at selected temperatures of the 1:1 (a) and 1:2 (c) ordered BCNO ceramics. The reciprocal temperature dependences of both *dc* conductivity ($\sigma_g$) and relaxation frequency ($f_m$) of the 1:1 (b) and 1:2 (d) ordered BCNO ceramics.

**Table Captions**

**Table 1:** List of sintering condition, relative density, average grain size, $\mathbf{A}_{1g}$ peak positions, and electrical parameters of the 1:1 and 1:2 ordered BCNO ceramics.

**Table 2:** Factor group analysis for the crystal structures addressed in this work ($\Gamma_T = \Gamma_{Ac} \oplus \Gamma_{Si} \oplus \Gamma_{IR} \oplus \Gamma_R$).

**Table 3:** Complex impedance, real part of electrical conductivity, real part of dielectric permittivity, and conductivity relaxation time for the parallel circuit models addressed in this work.

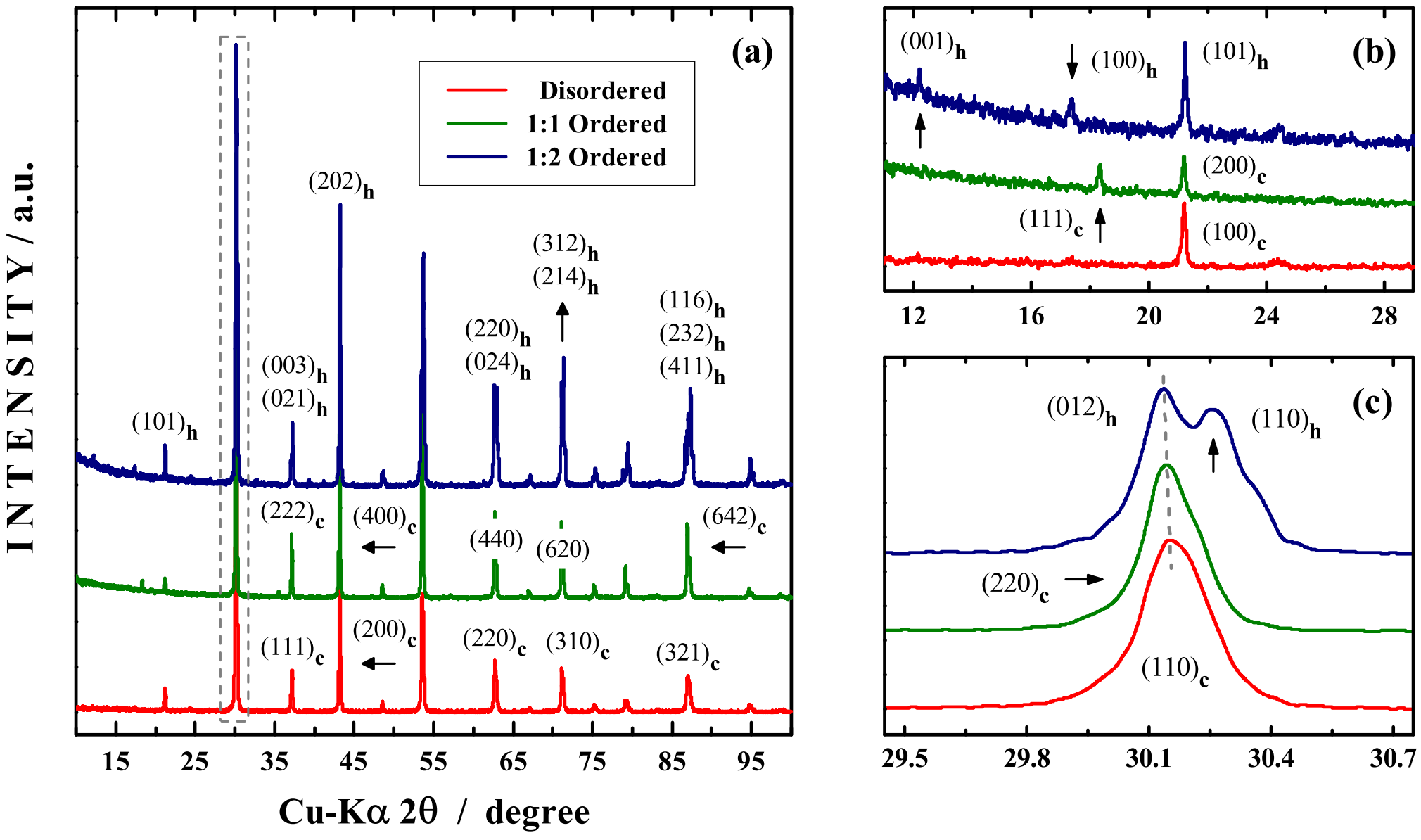

Figure 1.

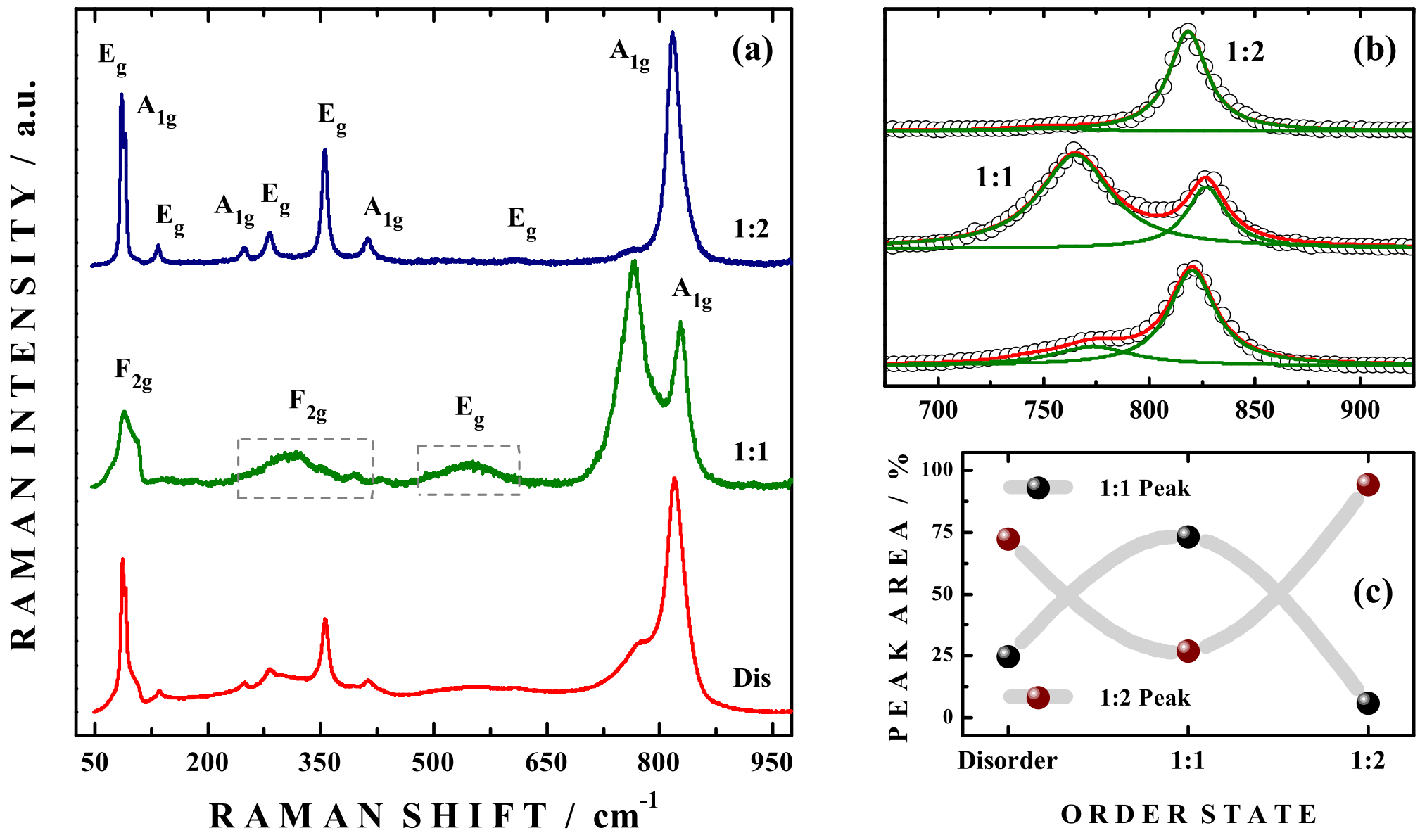

Figure 2.

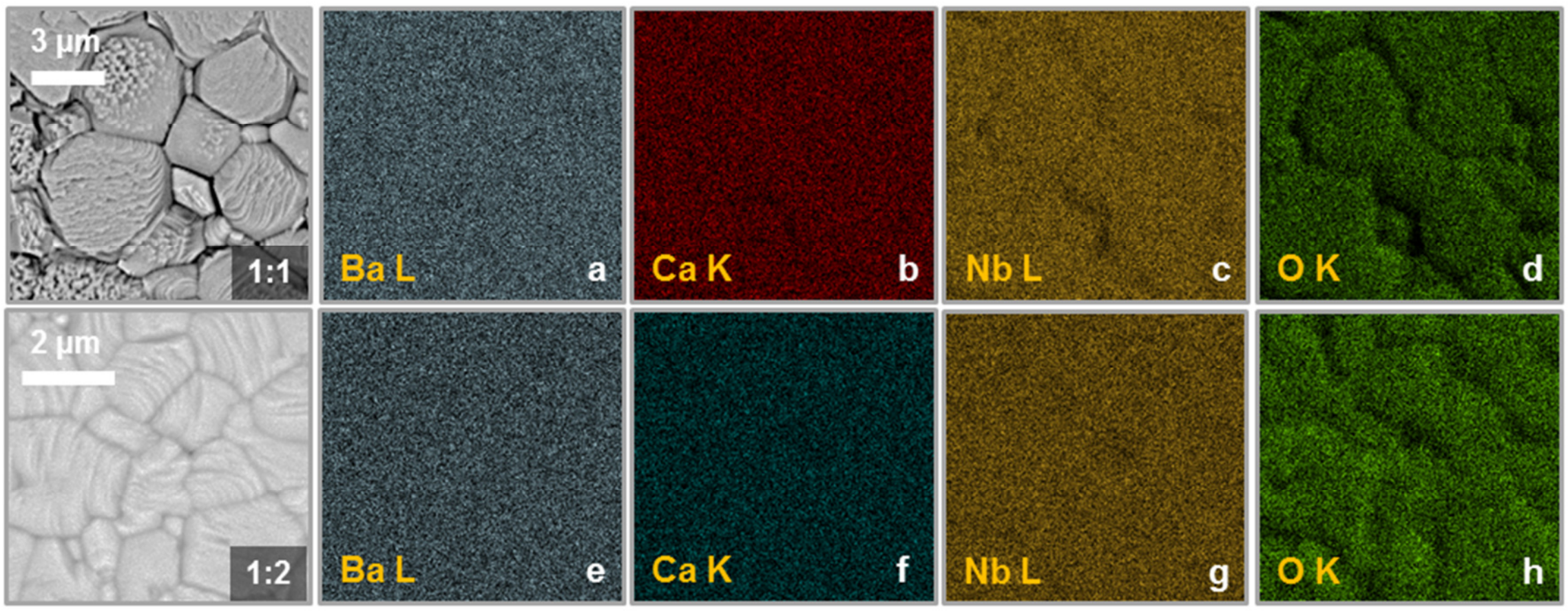

Figure 3.

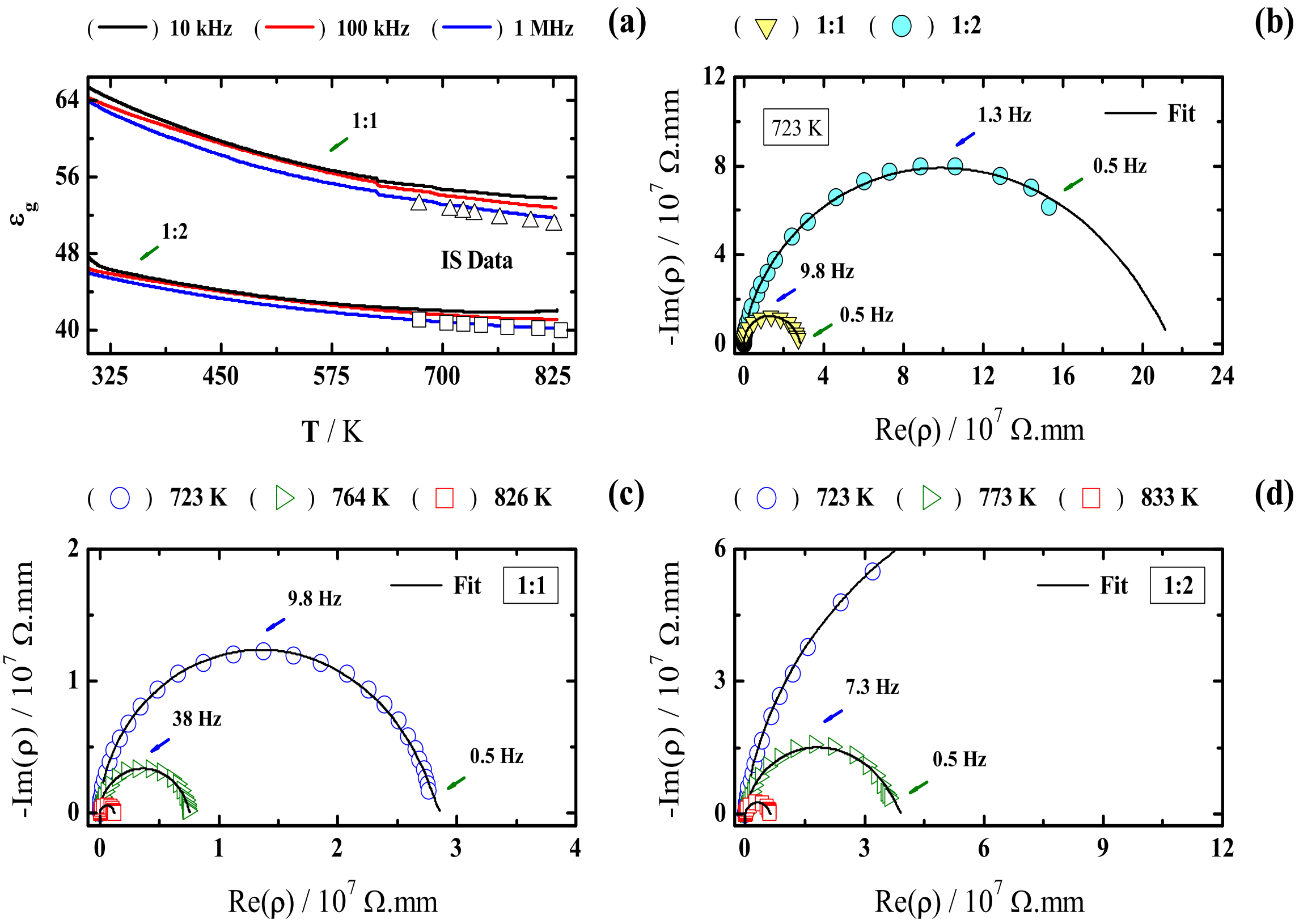

Figure 4.

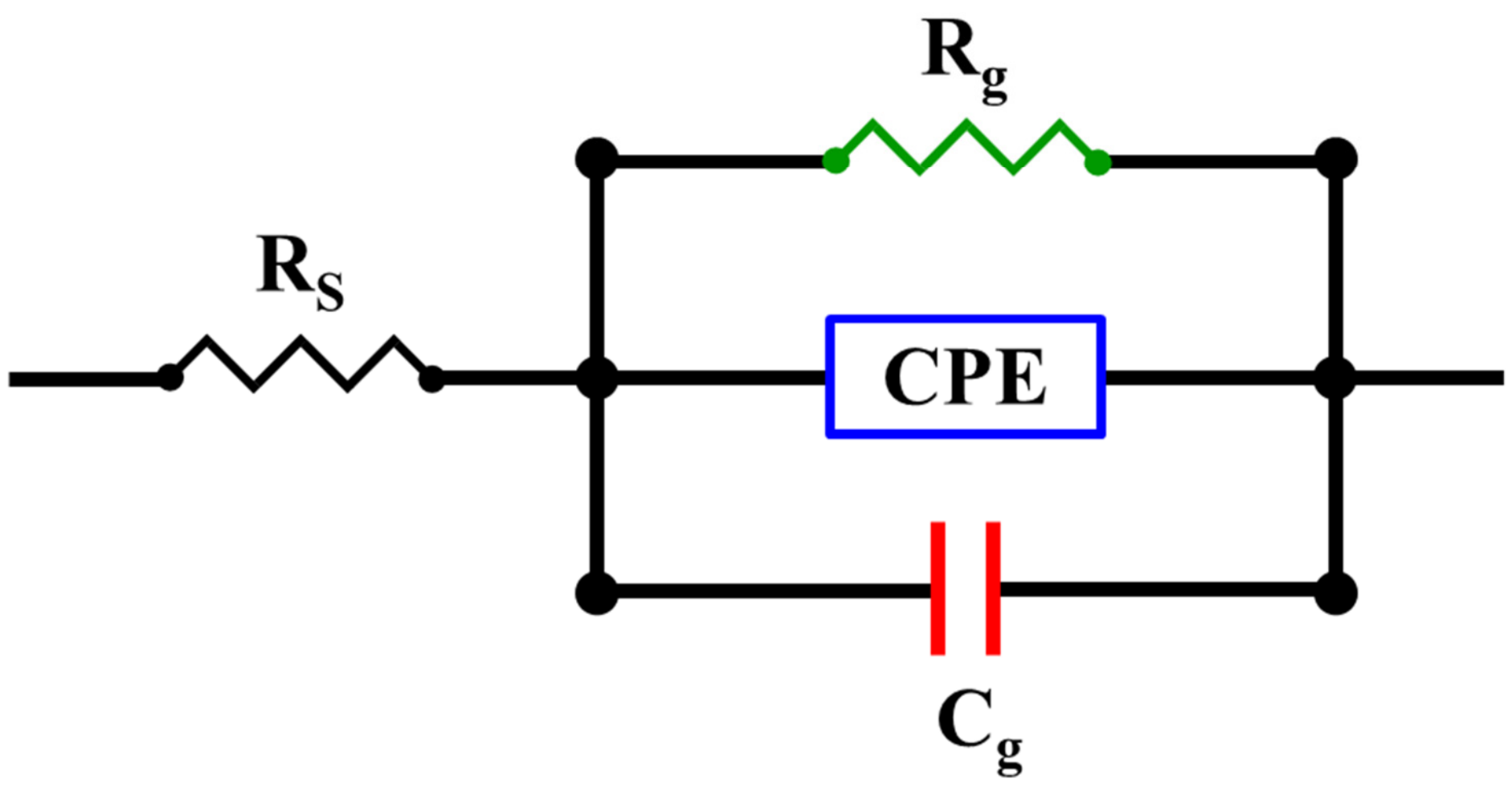

Figure 5.

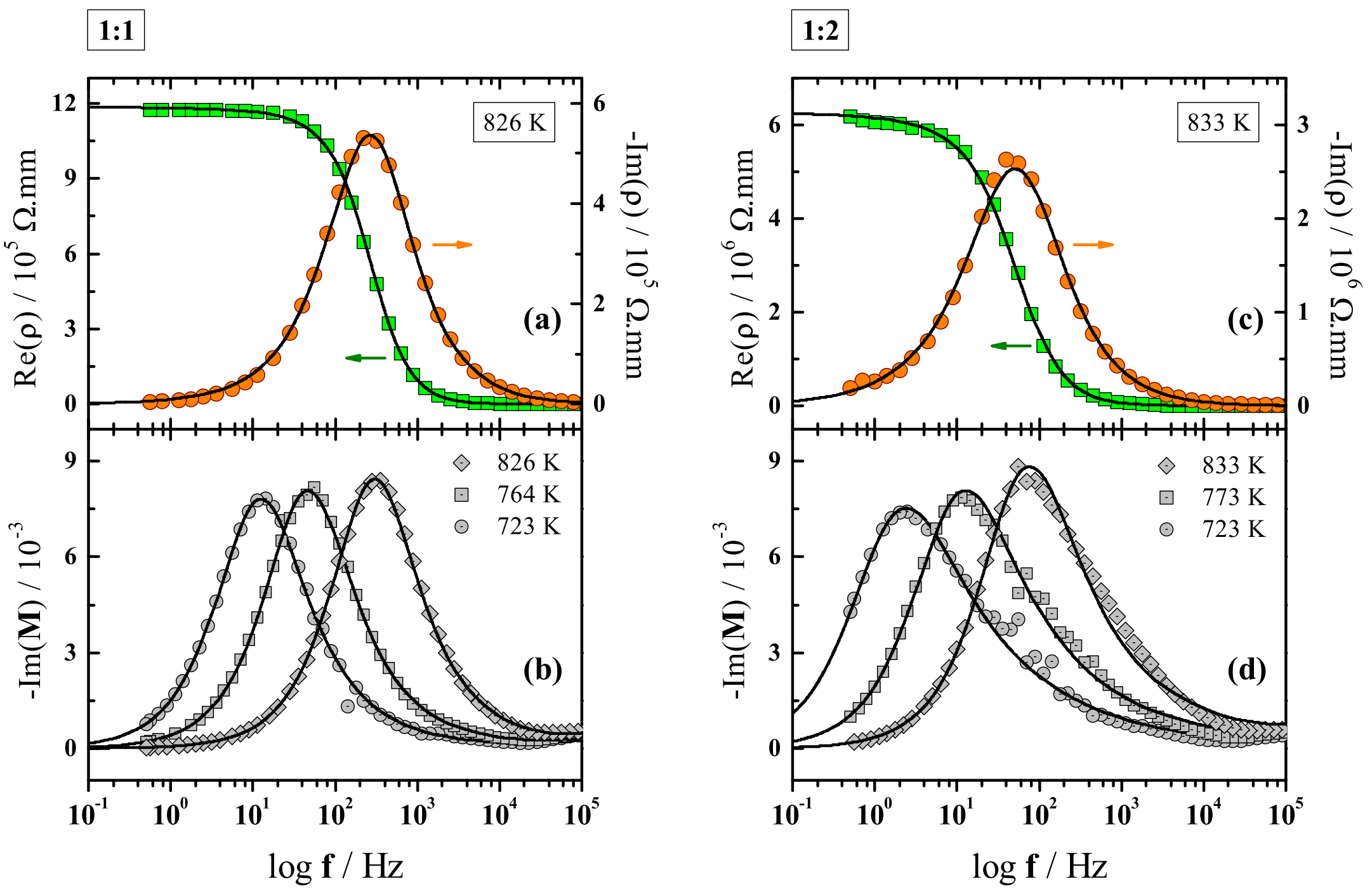

Figure 6.

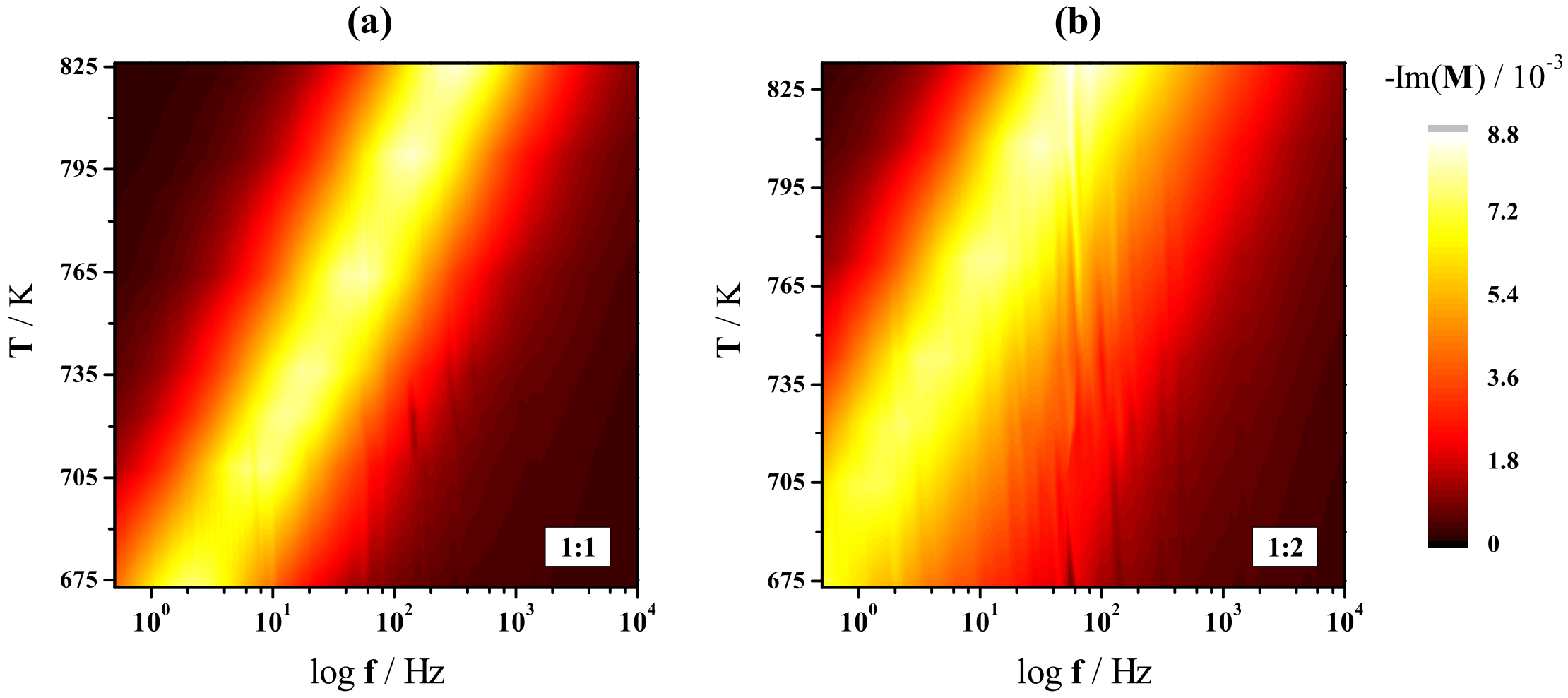

Figure 7.

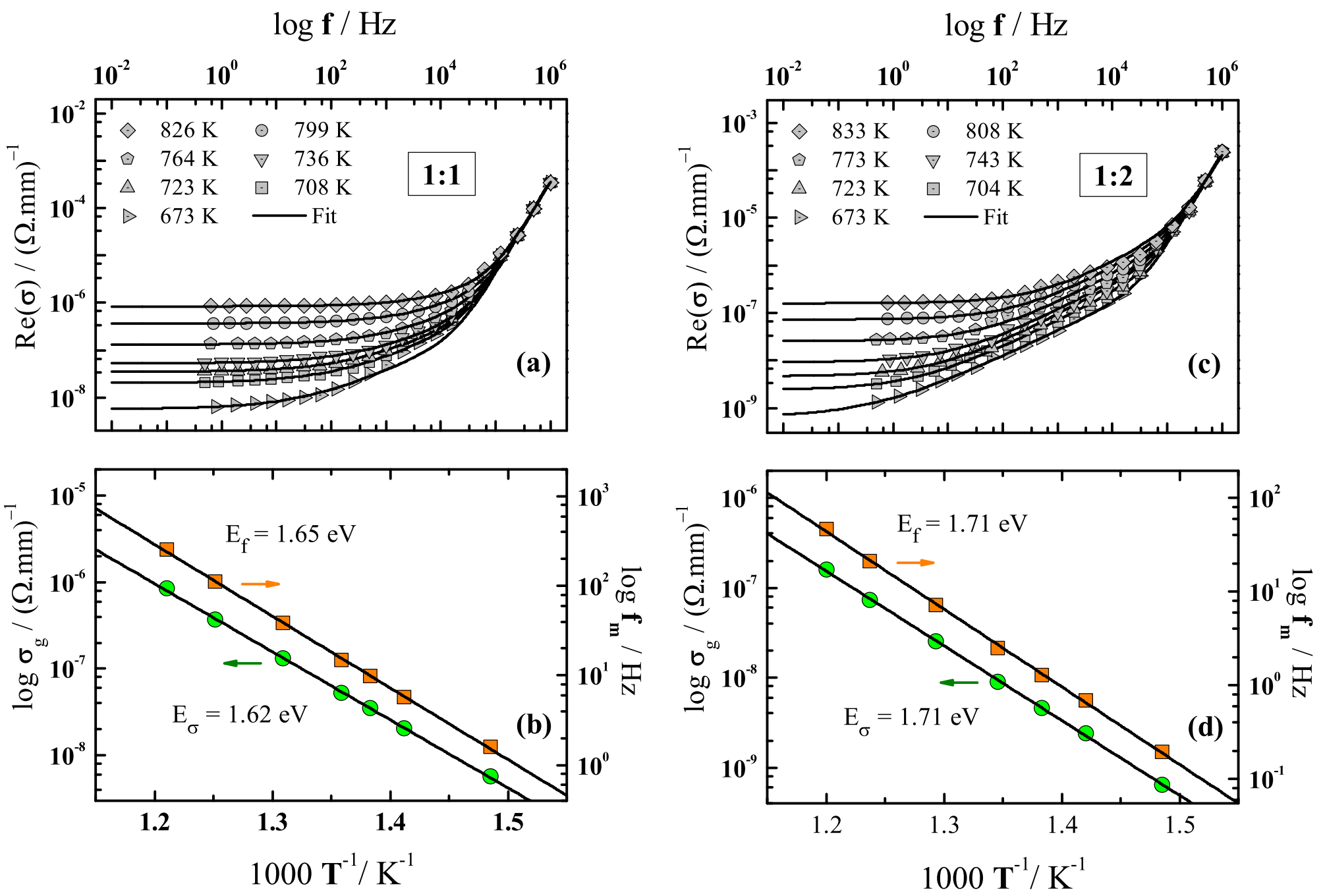

Figure 8.

Table 1

| | | |
|---|---|---|
| Long range order | 1:1 | 1:2 |
| Sintering condition | 1863 K for 2 h | 1773 K for 32 h |
| Relative density, % | 97 ± 1 | 92 ± 1 |
| Average grain size, µm | 2.8 ± 0.6 | 1.4 ± 0.5 |
| Space group | *Fm*-3*m* (225) | *P*-3*m*1 (164) |
| Crystallographic card | ICDD#49-0425 | ICSD#162758 |
| $\mathbf{A}_{1g}$ [1:1] peak position, $cm^{-1}$ | 765 | 750 |
| $\mathbf{A}_{1g}$ [1:2] peak position, $cm^{-1}$ | 827 | 818 |
| Geometrical factor λ, $mm^{-1}$ | 56.8 | 34.1 |
| $\sigma_g$, $10^{-9}$ $(\Omega.mm)^{-1}$ (at 723 K) | 34.9 | 4.63 |
| $\varepsilon_g$ (at 723 K) | 53.7 | 40.3 |
| α (at 723 K) | 0.547 | 0.611 |
| $R_g$, GΩ (at 723 K) | 1.63 | 7.37 |
| $C_g$, pF (at 723 K) | 8.4 | 10.5 |
| Activation energy $E_\sigma$, eV | 1.62 ± 0.02 | 1.71 ± 0.01 |
| $\tau_\varepsilon$, ppm.$K^{-1}$ | -225 | -143 |

Table 2

| Ion | Wyckoff site | Symmetry | Irreducible representation |
|---|---|---|---|
| **Disordered cubic:** $O_h^1$; *Pm*-3*m*; #221; $\boldsymbol{a^0a^0a^0}$ | | | |
| $Ba^{2+}$ | 1*b* | $O_h$ | $\mathbf{F}_{1u}$ |
| $Ca^{2+}/Nb^{5+}$ | 1*a* | $O_h$ | $\mathbf{F}_{1u}$ |
| $O^{2-}$ | 3*d* | $D_{4h}$ | $2\mathbf{F}_{1u} \oplus \mathbf{F}_{2u}$ |
| $\Gamma_T$ | $4\mathbf{F}_{1u} \oplus \mathbf{F}_{2u}$ | | |
| $\Gamma_{Ac}$ | $\mathbf{F}_{1u}$ | | |
| $\Gamma_{Si}$ | $\mathbf{F}_{2u}$ | | |
| $\Gamma_{IR}$ | $3\mathbf{F}_{1u}$ | | |
| $\Gamma_R$ | 0 | | |
| **1:1 Cubic:** $O_h^5$; *Fm*-3*m*; #225; $\boldsymbol{a^0a^0a^0}$ | | | |
| $Ba^{2+}$ | 8*c* | $T_d$ | $\mathbf{F}_{1u} \oplus \mathbf{F}_{2g}$ |
| $Ca^{2+}/Nb^{5+}$ | 4*a* | $O_h$ | $\mathbf{F}_{1u}$ |
| $Nb^{5+}$ | 4*b* | $O_h$ | $\mathbf{F}_{1u}$ |
| $O^{2-}$ | 24*e* | $C_{4v}$ | $\mathbf{A}_{1g} \oplus \mathbf{E}_g \oplus \mathbf{F}_{1g} \oplus \mathbf{F}_{2g} \oplus 2\mathbf{F}_{1u} \oplus \mathbf{F}_{2u}$ |
| $\Gamma_T$ | $\mathbf{A}_{1g} \oplus \mathbf{E}_g \oplus \mathbf{F}_{1g} \oplus 2\mathbf{F}_{2g} \oplus 5\mathbf{F}_{1u} \oplus \mathbf{F}_{2u}$ | | |
| $\Gamma_{Ac}$ | $\mathbf{F}_{1u}$ | | |
| $\Gamma_{Si}$ | $\mathbf{F}_{1g} \oplus \mathbf{F}_{2u}$ | | |
| $\Gamma_{IR}$ | $4\mathbf{F}_{1u}$ | | |
| $\Gamma_R$ | $\mathbf{A}_{1g} \oplus \mathbf{E}_g \oplus 2\mathbf{F}_{2g}$ | | |
| **1:2 Trigonal:** $D_{3d}^3$; *P*-3*m*1; #164; $\boldsymbol{a^0a^0a^0}$ | | | |
| $Ba^{2+}$ | 1*b* | $D_{3d}$ | $\mathbf{A}_{2u} \oplus \mathbf{E}_u$ |
| $Ba^{2+}$ | 2*d* | $C_{3v}$ | $\mathbf{A}_{1g} \oplus \mathbf{A}_{2u} \oplus \mathbf{E}_g \oplus \mathbf{E}_u$ |
| $Ca^{2+}$ | 1*a* | $D_{3d}$ | $\mathbf{A}_{2u} \oplus \mathbf{E}_u$ |
| $Nb^{5+}$ | 2*d* | $C_{3v}$ | $\mathbf{A}_{1g} \oplus \mathbf{A}_{2u} \oplus \mathbf{E}_g \oplus \mathbf{E}_u$ |
| $O^{2-}$ | 3*f* | $C_{2h}$ | $\mathbf{A}_{1u} \oplus 2\mathbf{A}_{2u} \oplus 3\mathbf{E}_u$ |
| $O^{2-}$ | 6*i* | $C_S$ | $2\mathbf{A}_{1g} \oplus \mathbf{A}_{1u} \oplus \mathbf{A}_{2g} \oplus 2\mathbf{A}_{2u} \oplus 3\mathbf{E}_g \oplus 3\mathbf{E}_u$ |
| $\Gamma_T$ | $4\mathbf{A}_{1g} \oplus \mathbf{A}_{2g} \oplus 5\mathbf{E}_g \oplus 2\mathbf{A}_{1u} \oplus 8\mathbf{A}_{2u} \oplus 10\mathbf{E}_u$ | | |
| $\Gamma_{Ac}$ | $\mathbf{A}_{2u} \oplus \mathbf{E}_u$ | | |
| $\Gamma_{Si}$ | $\mathbf{A}_{2g} \oplus 2\mathbf{A}_{1u}$ | | |
| $\Gamma_{IR}$ | $7\mathbf{A}_{2u} \oplus 9\mathbf{E}_u$ | | |
| $\Gamma_R$ | $4\mathbf{A}_{1g} \oplus 5\mathbf{E}_g$ | | |

Table 3

| **Circuit model** | **Z(ω)** | **Re(σ)** | **Re(ε)** | $\boldsymbol{\tau_\sigma}$ |
|---|---|---|---|---|
| $R_g \parallel C_g$ | $\dfrac{R_g}{1+j\omega R_g C_g}$ | $\sigma_g$ | $\varepsilon_g$ | $R_g C_g$ |
| $R_g \parallel CPE$ | $\dfrac{R_g}{1+(j\omega)^{\alpha} R_g Q}$ | $\sigma_g + \lambda Q \cos\left(\dfrac{\alpha\pi}{2}\right)\omega^{\alpha}$ | $\lambda Q \sin\left(\dfrac{\alpha\pi}{2}\right)\omega^{\alpha-1}$ | $(R_g Q)^{\frac{1}{\alpha}}$ |
| $R_g \parallel CPE \parallel C_g$ | $\dfrac{R_g}{1+j\omega R_g C_g + (j\omega)^{\alpha} R_g Q}$ | $\sigma_g + \lambda Q \cos\left(\dfrac{\alpha\pi}{2}\right)\omega^{\alpha}$ | $\varepsilon_g + \lambda Q \sin\left(\dfrac{\alpha\pi}{2}\right)\omega^{\alpha-1}$ | $\approx R_g C_g$ |